\documentclass[aps,prl,twocolumn,amsmath,amssymb,superscriptaddress,nofootinbib,showpacs]{revtex4-1}
\usepackage{aas_macros}
\usepackage{graphicx}
\usepackage{bm}
\usepackage{wasysym}

\usepackage[colorlinks=true]{hyperref}

\newcommand{\bs}{\bar s}

\DeclareTextSymbol{\degre}{T1}{6}
\DeclareTextSymbol{\degre}{OT1}{23}

\def\bs{\bar s}

\def\bx{{\bm x}}
\def\bk{{\bm k}}

\def\bn{{\bm n}}
\def\bp{{\bm p}}

\def\T{{\mathcal T}}

\begin{document}

\title{Lorentz symmetry and Very Long Baseline Interferometry}

\author{C.~Le~Poncin-Lafitte}
\email{christophe.leponcin@obspm.fr}
\affiliation{SYRTE, Observatoire de Paris, PSL Research University, CNRS, Sorbonne Universit\'es, UPMC Univ. Paris 06, LNE, 61 avenue de l'Observatoire, 75014 Paris, France}

\author{A.~Hees}
\email{ahees@astro.ucla.edu}
\affiliation{Department of Mathematics, Rhodes University, 6140 Grahamstown, South Africa}
\affiliation{Department of Physics and Astronomy, University of California, Los Angeles, CA 90095, USA}

\author{S.~Lambert}
\email{sebastien.lambert@obspm.fr}
\affiliation{SYRTE, Observatoire de Paris, PSL Research University, CNRS, Sorbonne Universit\'es, UPMC Univ. Paris 06, LNE, 61 avenue de l'Observatoire, 75014 Paris, France}

\date{\today}


\begin{abstract}
Lorentz symmetry violations can be described by an effective field theory framework that contains both General Relativity and the Standard Model of particle physics called the Standard-Model extension (SME). Recently, post-fit analysis of Gravity Probe B and binary pulsars lead to an upper limit at the $10^{-4}$ level on the time-time coefficient $\bar s^{TT}$ of the pure-gravity sector of the minimal SME. In this work, we derive the observable of Very Long Baseline Interferometry (VLBI) in SME and then we implement it into a real data analysis code of geodetic VLBI observations. Analyzing all available observations recorded since 1979, we compare estimates of $\bar s^{TT}$ and errors obtained with various analysis schemes, including global estimations over several time spans and with various Sun elongation cut-off angles, and with analysis of radio source coordinate time series. We obtain a constraint on $\bar s^{TT}=(-5\pm 8)\times 10^{-5}$, directly fitted to the observations and improving by a factor 5 previous post-fit analysis estimates.
\end{abstract}

\maketitle

\par Historically, the measurement of the bending of light due to the gravitational mass of the Sun is one of the most important and precise test of General Relativity (GR). Within the Parameterized Post-Newtonian (PPN) formalism \cite{1993tegp.book.....W,*will:2014la}, this effect has been constrained by Very Long Baseline Interferometry (VLBI) observations \cite{1995PhRvL..75.1439L,2009A&A...499..331L,*2011A&A...529A..70L}, space astrometry with Hipparcos \cite{1997ESASP.402...49F} and the Cassini radioscience experiment \cite{2003Natur.425..374B}, the latter being the most stringent constraint on the PPN $\gamma$ parameter. 

\par The SME framework has been developed to be an extensive formalism that allows a systematic description of  Lorentz symmetry violations in all sectors of physics, including gravity~\cite{1997PhRvD..55.6760C,1998PhRvD..58k6002C,2004PhRvD..69j5009K}. If the motivations came first from string theory~\cite{1989PhRvD..39..683K,1989PhRvD..40.1886K} which can possibly produce Lorentz violations, this statement appears also in loop quantum gravity, non commutative field theory and others~\cite{2005LRR.....8....5M,2014RPPh...77f2901T}. 

\par A hypothetical Lorentz violation in the gravitational sector naturally leads to an expansion at the level of the action~\cite{2004PhRvD..69j5009K,2006PhRvD..74d5001B} which in the minimal SME writes
\begin{eqnarray}
	S_\textrm{grav}&=&\int d^4x\frac{\sqrt{-g}}{16\pi G}\left(R-uR+s^{\mu\nu}R^T_{\mu\nu}+t^{\alpha\beta\mu\nu}C_{\alpha\beta\mu\nu}\right) \nonumber\\
	&&\qquad + S'[s^{\mu\nu},t^{\alpha\beta\mu\nu},g_{\mu\nu}]\, ,\label{eq:action}
\end{eqnarray}
with $G$  the gravitational constant, $g$  the determinant of the space-time metric $g_{\mu\nu}$, $R$ the Ricci scalar, $R^T_{\mu\nu}$ the trace-free Ricci tensor, $C_{\alpha\beta\mu\nu}$ the Weyl tensor and $u$, $s^{\mu\nu}$ and $t^{\alpha\beta\mu\nu}$ the Lorentz violating fields.  To avoid conflicts with the underlying Riemann geometry, we assume spontaneous symmetry breaking so that the Lorentz violating coefficients need to be considered as dynamical fields~\cite{2006PhRvD..74d5001B}. The last part of the action $S'$ contains the dynamical terms governing the evolution of the SME coefficients. In the linearized gravity limit, the metric depends only on $\bar u$ and $\bar{s}^{\mu\nu}$ which are the vacuum expectation value of $u$ and $s^{\mu\nu}$~\cite{2006PhRvD..74d5001B}. The coefficient $\bar u$ is unobservable since it can be absorbed in a rescaling of the gravitational constant. The so obtained post-Newtonian metric differs from the one introduced in the PPN formalism~\cite{2006PhRvD..74d5001B}. In addition to Lorentz symmetry violations in the pure-gravity sector, violations of Lorentz symmetry can also arise from gravity-matter couplings~\cite{2011PhRvD..83a6013K}, but we do not consider them in this work. Hence SME is an effective field theory making possible confrontations of fundamental theories and experiments. Indeed, since the last decade, several studies aimed to find upper limit on SME coefficients by searching possible signals in post-fit residuals of experiments. This was done for pure-gravity SME coefficients with Lunar Laser Ranging \cite{2007PhRvL..99x1103B}, atom interferometry~\cite{muller:2008kx,*chung:2009uq}, Gravity probe B \cite{2013PhRvD..88j2001B}, binary pulsars \cite{2014PhRvL.112k1103S,*2014PhRvD..90l2009S}, Solar System planetary motions \cite{iorio:2012zr,2015PhRvD..92f4049H}, cosmic ray observations~\cite{kostelecky:2015db} or event very recently with gravitational waves detection~\cite{kostelecky:2016nx}. However, all these works are post-fit analysis based originally on pure GR and consequently their approach is not fully satisfactory in the sense that correlations in the determination of SME coefficients and other global parameters (masses, position and velocity\ldots) can not be assessed. Then in the best case, a simple modeling of extra terms containing SME coefficients are least square fitted in the residuals of the experiment. In a more correct approach, SME modeling must be included in the complete data analysis and its coefficients must be determined as global parameters. It is exactly what we present here in the case of VLBI observations.

\par VLBI is a geometric technique which measures the time difference in the arrival of a radio wavefront emitted by a distant radio source (typically a quasar) between at least two Earth-based radio telescopes, with a precision of a few picoseconds. Knowing the group delay and the angular separation between the baseline between the antennas of the telescopes and the line of sight of the observation, the distance between the telescopes can be determined and consequently VLBI tracks the orientation of the Earth in an inertial reference frame provided by the very distant quasars, determining accurate terrestrial and celestial reference frames.

\par Let us write the VLBI group delay in the International Celestial Reference Frame (ICRF) as defined by the International Astronomical Union (IAU)~\cite{2003AJ....126.2687S} with coordinates $(x^\mu)=(x^T,\bx )$, where $x^T=ct$, $t$ being a time coordinate, and $\bx=(x^I)$ is the spatial position. We consider a quasar as source with as coordinates of the emission event $(t_e,\bx_e)$. This signal is received by two different VLBI stations at events $(t_1,\bx_1)$ and $(t_2,\bx_2)$, respectively. Using the same notations as in \cite{1983Ap&SS..94..233F}, we introduce three units vectors
\begin{equation}
	\bk=\frac{\bx_e}{|\bx_e|}\, , \quad	\bn_{ij}\equiv\frac{\bx_{ij}}{r_{ij}}=\frac{\bx_j-\bx_i}{|\bx_{ij}|}\, , \quad \textrm{and} \quad
	\bn_i=\frac{\bx_i}{|\bx_i|}\, .
\end{equation}
\par We denote by $t_r-t_e=\T(\bx_e,t_e,\bx_r)$ the coordinate propagation time of a photon between an emission event whose coordinates are given by $(t_e,\bx_e)$ and a reception event whose coordinates are given by $(t_r,\bx_r)$. We deduce simply the VLBI group delay $\Delta\tau$ from
\begin{equation}
	\Delta \tau(\bx_e,t_e,\bx_1,\bx_2)=\T(\bx_e,t_e,\bx_2)-\T(\bx_e,t_e,\bx_1)\, .
\end{equation}
For the observation of a quasar, we then use the limit $r_e\equiv\vert\bx_e\vert\rightarrow \infty$ and the VLBI time delay is given by
\begin{equation}\label{eq:deltat}
	\Delta \tau(\bk,\bx_1,\bx_2)=\lim\limits_{r_e\rightarrow \infty}\left\lbrack\T(\bx_e,t_e,\bx_2)-\T(\bx_e,t_e,\bx_1)\right\rbrack\, .
\end{equation}

\par The coordinate propagation time can be computed from the linearized SME metric from~\cite{2006PhRvD..74d5001B,tso:2011uq} using the time transfer functions formalism~\cite{2004CQGra..21.4463L,*teyssandier:2008nx,*le-poncin-lafitte:2008fk,*hees:2014fk,*hees:2014nr}. In SME, it has been computed in \cite{2009PhRvD..80d4004B} (see Eq.~(24)) for the pure gravity sector and is given by
\begin{align}      
	\T& (\bx_e,t_e,\bx_r)=\frac{r_{er}}{c}+\frac{GM}{c^3} \Big(\bs^{TJ}p_{er}^J - \bs^{JK}n_{er}^J p_{er}^K\Big)\frac{r_e-r_r}{r_er_r}\nonumber \\
	&+2\frac{GM}{c^3}\left[1+\bs^{TT}-\bs^{TJ}n_{er}^J\right] \ln \frac{r_e-\bn_{er}.\bx_e}{r_r-\bn_{er}.\bx_r}  \label{eq:delay} \\
     &+\frac{GM}{c^3} \Big\lbrack\bs^{TJ} n_{er}^J +\bs^{JK}\hat p_{er}^J\hat p_{er}^K-\bs^{TT}\Big\rbrack\left(\bn_r . \bn_{er}-\bn_e . \bn_{er}\right)\nonumber
 \end{align}
  where the terms $a_1$ and $a_2$ from \cite{2009PhRvD..80d4004B} are taken as unity (which corresponds to using the harmonic gauge, which is the one used for VLBI data reduction) and where 
\begin{eqnarray}
\bp_{er}&=&\bn_{er}\times(\bx_r\times \bn_{er})=\bx_r-(\bn_{er}.\bx_r)\bn_{er}\nonumber\\
&=&\bn_{er}\times(\bx_e\times \bn_{er})=\bx_e-(\bn_{er}.\bx_e)\bn_{er},
\end{eqnarray}
and where $\hat \bp_{er}=\frac{\bp_{er}}{|\bp_{er}|}$.
  
\par We can now give the expression of the group delay between two VLBI stations. We are using the assumptions that the source is located at infinity ($r_e \rightarrow \infty$). We need to introduce (\ref{eq:delay}) into (\ref{eq:deltat}), which leads to
\begin{widetext}
\begin{eqnarray}
	\Delta \tau_{(\textrm{grav})}(\bk,\bx_1,\bx_2)&=&2\frac{GM}{c^3}\left[1+\bs^{TT}+\bs^{TJ} k^J\right] \ln  \frac{r_1+\bk.\bx_1}{r_2+\bk.\bx_2}  \nonumber\\
	&&+   \frac{GM}{c^3} \Big[\bs^{TT} -\bs^{JK}k^Jk^K\Big]\left(\bn_2 . \bk-\bn_1 . \bk \right) \label{eq:full_VLBI} \\
	&&+\frac{GM}{c^3}\Big[ \bs^{TJ}+\bs^{JK}k^K \Big](n_2^J-n_1^J) +\frac{GM}{c^3}\Big[ \bs^{JK}\hat p_1^J\hat p_1^K (\bn_1.\bk-1)-\bs^{JK}\hat p_2^J\hat p_2^K (\bn_2.\bk-1) \Big]\, ,\nonumber
\end{eqnarray}
\end{widetext}
where the subscript (grav) refers to the gravitational part of the group delay and where 
\begin{equation}
	\bp_i=\bm k \times (\bx_i \times \bm k)=\bx_i-(\bm k\ .\ \bm x_i)\bm k \, ,
\end{equation}
and $\hat \bp_{i}=\frac{\bp_{i}}{|\bp_{i}|}$.
 Moreover, a simplified formula can be used for practical utilisation considering a typical accuracy of a VLBI observation of the order of 10 ps and that $GM/c^3\sim 5\times 10^{-6}$ s. Since the coefficients $\bs^{TJ}$ are already constrained and are smaller than $\sim 10^{-7}$ \cite{2011RvMP...83...11K,2015PhRvD..92f4049H}, all terms $GM/c^3\bs^{TJ}$ are too small to be detected and can be neglected. The coefficients $\bs^{IJ}$ with $I\neq J$ are also constrained by previous studies and are smaller than $10^{-10}$~\cite{2011RvMP...83...11K,2014PhRvL.112k1103S,2015PhRvD..92f4049H}. Therefore, we can also neglect terms that are proportional to $GM/c^3\bs^{IJ}$ with $I\neq J$. Finally, since we know that $|\bs^{XX}-\bs^{YY}|<10^{-10}$ and $|\bs^{XX}+\bs^{YY}-2\bs^{ZZ}|<10^{-10}$~\cite{2011RvMP...83...11K}, we can safely say that at the level of accuracy required $\bs^{XX}\approx\bs^{YY}\approx\bs^{ZZ}$ in Eq.~(\ref{eq:full_VLBI}). Under these assumptions and using the fact that $\bs^{\mu\nu}$ is traceless, the VLBI group delay can be written
\begin{eqnarray}
	\Delta \tau_{(\textrm{grav})}&=&2\frac{GM}{c^3}(1+\bs^{TT}) \ln  \frac{r_1+\bk.\bx_1}{r_2+\bk.\bx_2} \nonumber\\
	&& \qquad\qquad+\frac{2}{3}\frac{GM}{c^3} \bs^{TT}\left(\bn_2 . \bk-\bn_1 . \bk \right)\, .
\end{eqnarray}
It is important to notice that the bare $GM$ parameter appearing in the post-Newtonian metric do not correspond to the observed $\widetilde{GM}$ parameter measured with orbital dynamics (using planetary motion for the Sun). There is a rescaling between the two parameters given by $\widetilde{GM}=GM\left(1+5/3\bar s^{TT}\right)$ (see Sec.~IV of~\cite{2013PhRvD..88j2001B} or~\cite{2006PhRvD..74d5001B,2015PhRvD..92f4049H}). Using the observed mass parameter leads to 
\begin{eqnarray}
	\Delta \tau_{(\textrm{grav})}&=&2\frac{\widetilde{GM}}{c^3}(1-\frac{2}{3}\bs^{TT}) \ln  \frac{r_1+\bk.\bx_1}{r_2+\bk.\bx_2} \nonumber\\
	&& \qquad\qquad+\frac{2}{3}\frac{\widetilde{GM}}{c^3} \bs^{TT}\left(\bn_2 . \bk-\bn_1 . \bk \right)\, .\label{eq:vlbi_stt}
\end{eqnarray}
This last formula is the one used to fit the $\bs^{TT}$ coefficient using VLBI observations.

\par
\begin{figure}[htbp]
\begin{center}
\includegraphics[width=8cm]{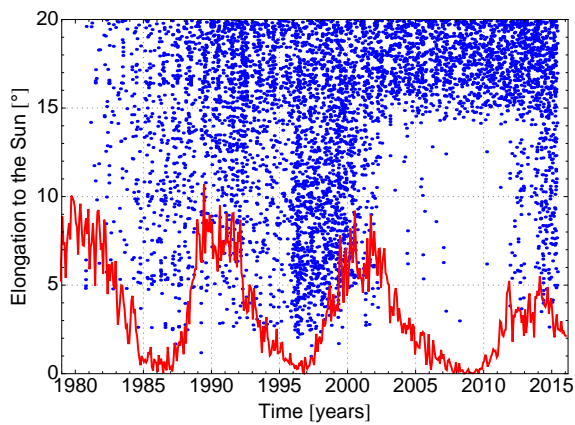}
\end{center}
\caption{Observational history of the sources at less than $20^{\circ}$ to the Sun (blue dots) and Sun spot number (red curve, rescaled to fit in the plot;~\cite{2007AdSpR..40..919C}).}
\label{figprog}
\end{figure}

 From August 1979 to mid-2015, almost 6000 VLBI 24-hr sessions (correspondingly 10 million delays) have been scheduled for primary goal of monitoring the Earth's rotation and determining reference frames. The International VLBI Service for Geodesy and Astrometry (IVS)~\cite{2012JGeo...61...68S}\footnote{The IVS operates regular geodetic VLBI since 1998.} imposed a minimal distance to the Sun of $15^{\circ}$ after 2002 in order to avoid potential degradation of geodetic products due to radio wave crossing of the Solar corona. This limit was recently removed (Fig.~\ref{figprog}). 
\par In our analysis, all VLBI delays were corrected from delay due to the radio wave crossing of dispersive regions in the signal propagation path in a preliminary step that made use of 2~GHz and 8~GHz recordings. Then, we only used the 8~GHz delays to fit the parameters listed hereafter. We used the Calc/Solve geodetic VLBI analysis software developed at NASA Goddard Space Flight Center, in which the astrometric modelling of VLBI time delay is compliant with the latest standards of the International Earth Rotation and Reference Systems Service (IERS)~\cite{2010ITN....36....1P}. We added the partial derivative of the VLBI delay with respect to $\bar s^{TT}$ from Eq.~(\ref{eq:vlbi_stt}) to the software package using the USERPART module of Calc/Solve.

\par
\begin{figure}[htbp]
\begin{center}
\includegraphics[width=8cm]{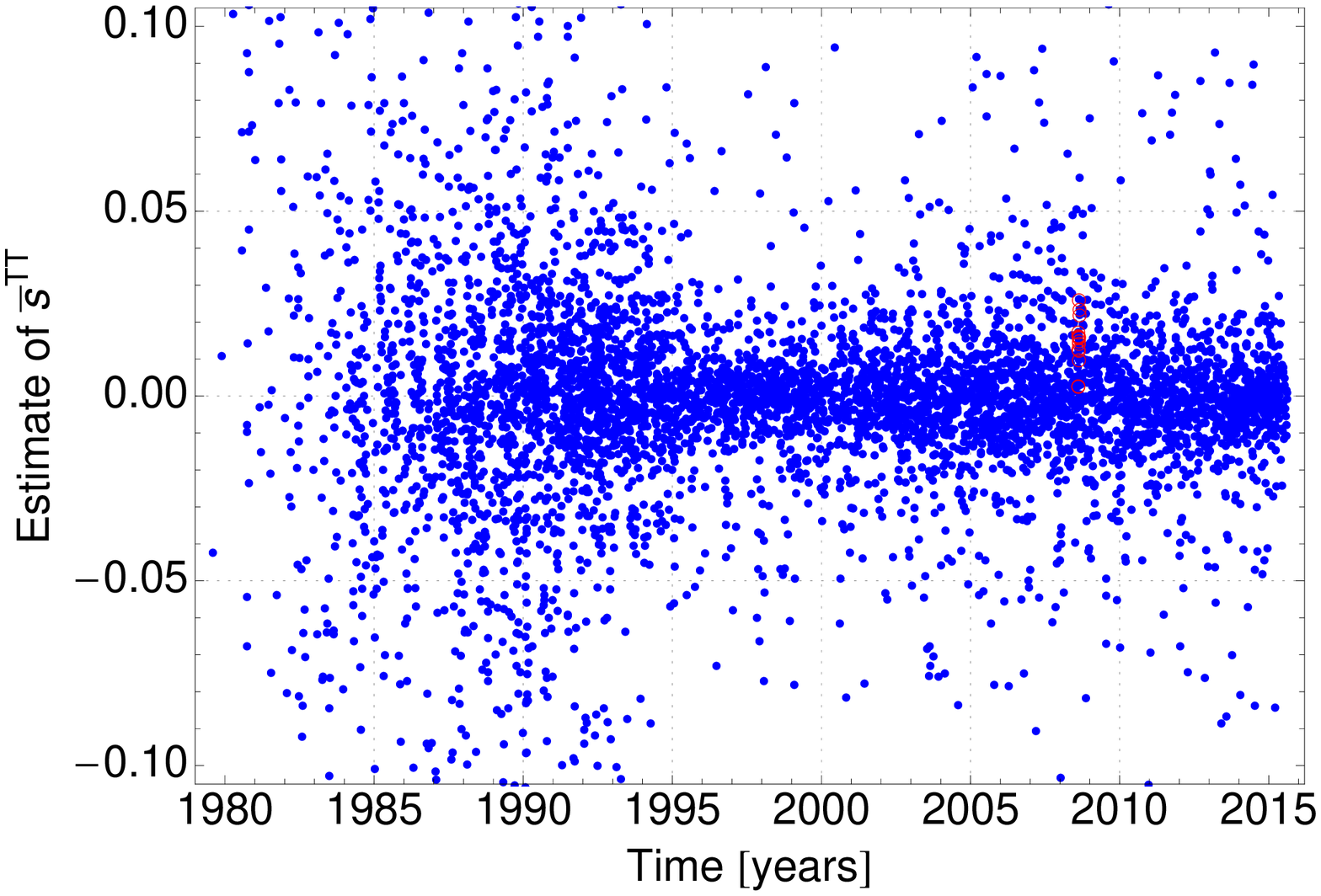}
\includegraphics[width=8cm]{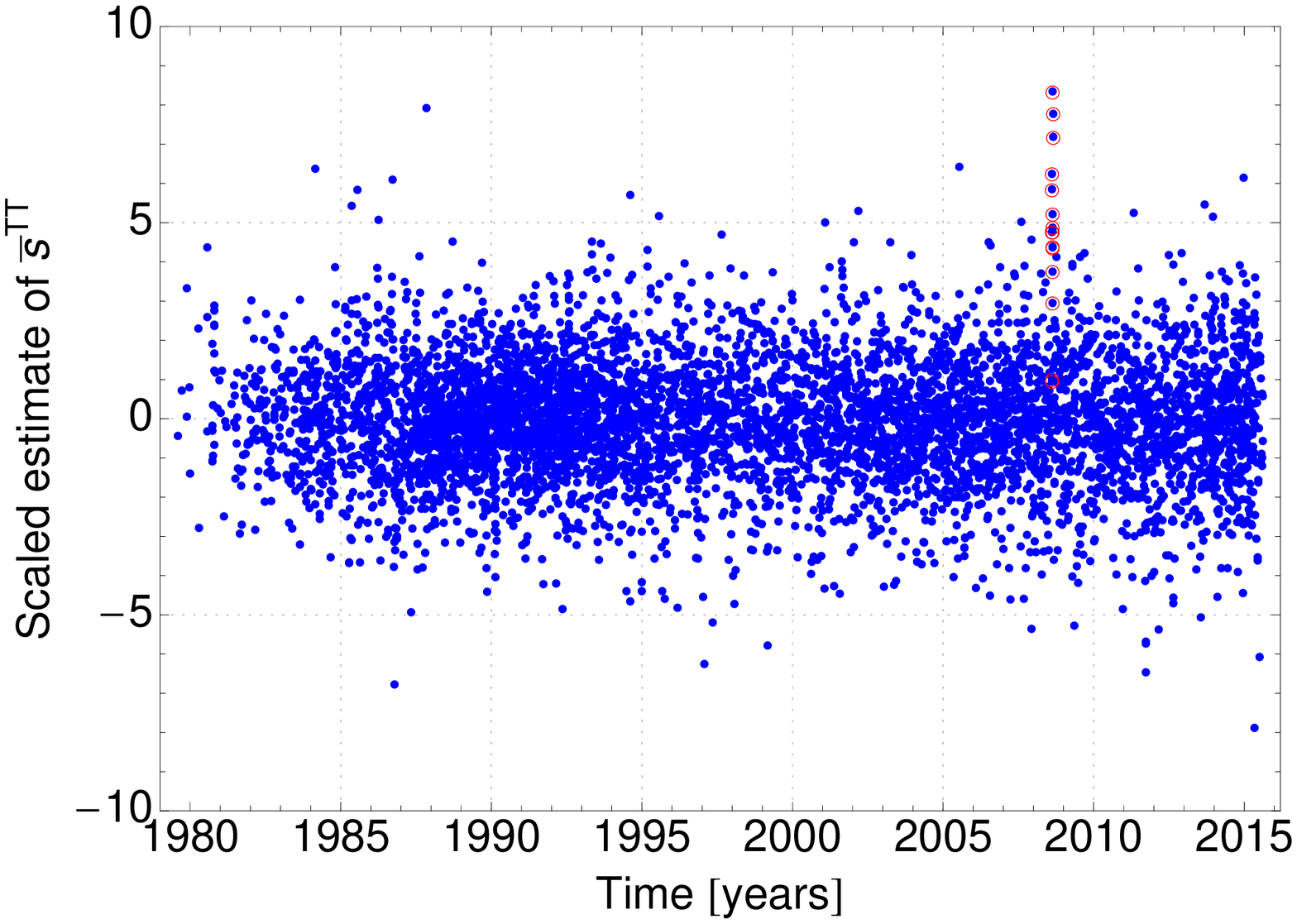}
\end{center}
\caption{Session-wise estimates of $\bar s^{TT}$ (Top) and $\bar s^{TT}$ scaled by its error (Bottom) for 5895 sessions (blue dots). The red circles highlight the 26 CONT08 sessions.}
\label{figcont}
\end{figure}

 We ran a first solution in which we estimated $\bar s^{TT}$, all source and station coordinates and all five Earth orientation parameters once per session. A priori zenith delays were determined from local pressure values \cite{1972GMS....15..247S}, which were then mapped to the elevation of the observation using the Vienna mapping function~\cite{2006JGRB..111.2406B}. Wet zenith delays and clock drifts were estimated at intervals of ten and thirty minutes, respectively. Troposphere gradients were estimated at intervals of 6~hours. Suitable loose constraints were applied to source and station coordinates to avoid global rotation of the celestial frame and global rotation and translation of the terrestrial frame. Sites undergoing strong nonlinear motions due to, e.g., post-seismic relaxation, were excluded from the constraint. This preliminary solution allowed us to identify a half-dozen of sessions with abnormally high postfit rms (generally higher than 1~ns). The distribution $\bar s^{TT}$ scaled by its error also reveals a few points clearly lying outside the distribution (see Fig.~\ref{figcont}). These data corresponds to the 26 sessions of the CONT08 campaign (August 2008), representing 1.1\% of the dataset. Without the CONT08 sessions, we obtained $\bar s^{TT}=(-5\pm11)\times10^{-5}$. Keeping the CONT08 sessions moves the mean value to~$7\times10^{-5}$.

\par A spectral analysis of the time series revealed no significant peak. We computed $\bar s^{TT}$ over 1000 random subsets containing three quarters of the 5895 sessions to check the stability of the mean value. $\bar s^{TT}$ stays around 0 within $8\times10^{-5}$. We also addressed the sensitivity to Solar activity. To do so, we used the Sun spot number (SSN) monthly data to separate VLBI sessions into two groups: each group contains sessions occurring when the SSN is higher or lower than its median value computed over our observational time span, that is 2947 sessions in each group. We obtained $\bar s^{TT}=(3\pm16)\times10^{-5}$ for the high activity periods, and $\bar s^{TT}=(-12\pm15)\times10^{-5}$ for low activity period, giving no clue on the influence of Solar activity.

\par We turned to a global solution in which we estimated $\bar s^{TT}$ as a global parameter together with radio source coordinates. Station coordinates were left as session parameters. Constraints remained unchanged. We obtained $\bar s^{TT}=(-5\pm8)\times10^{-5}$, with a global postfit rms of 28~ps and a $\chi^2$ per degree of freedom of 1.15. Correlations between radio source coordinates and $\bar s^{TT}$ remain lower than 0.02. The global estimate is consistent with the mean value obtained with the session-wise solution with a slightly lower error.

\par In this letter, we have presented a test of Lorentz symmetry performed using 36 years of VLBI data. Contrarily to previous studies of Lorentz symmetry in the gravity sector, our work is not based on a post-fit analysis on residuals obtained after a GR analysis but rather on a full SME modelling in the VLBI data reduction process. Our analysis leads to a constraint on the $\bar s^{TT}$ coefficient at the level of $10^{-5}$. This coefficient is particularly important since it controls the speed of gravity in the SME framework~\cite{kostelecky:2016nx}. Our result improves the best current constraint on this coefficient~\cite{2013PhRvD..88j2001B,2014PhRvL.112k1103S,*2014PhRvD..90l2009S} by a factor of~five. In the future, the accumulation of VLBI data in the framework of the permanent geodetic monitoring program let us expect improvements of this constraint as well as extended tests.

\par {\it Acknowledgments.} The authors thank Q. Bailey for useful comments on a preliminary version of this manuscript. C.L.P.L. is grateful for the financial support of CNRS/GRAM and Axe Gphys of Paris Observatory Scientific Council. This study could not have been carried out without the work of the International VLBI Service for Geodesy and Astrometry (IVS) community that coordinates observations and correlates and stores geodetic VLBI data.

\bibliographystyle{apsrev4-1}
\bibliography{SME_VLBI}
\end{document}